\documentclass{pic2012}
\def\runauthor{Matthew Wing}
\def\shorttitle{Measurements of deep inelastic scattering at HERA}
\begin{document}

\title{MEASUREMENTS OF DEEP INELASTIC \\
SCATTERING AT HERA
}

\author{MATTHEW WING}

\address{Department of Physics and Astronomy, UCL\\
Gower Street, London WC1E 6BT, UK\\
E-mail: m.wing@ucl.ac.uk }

\maketitle

\abstracts{
After fifteen years of running and a further five years of analysis, the final inclusive deep inelastic scattering cross sections from 
H1 and ZEUS have been published.  Measurements of neutral current and charged current processes in $ep$ collisions at HERA 
are presented.  These provide us with the most valuable information on the structure of the proton, which tells us about the 
fundamental structure of matter and is essential for understanding processes at proton colliders such as the Large Hadron 
Collider.  The measurements also demonstrate the chiral structure of the weak force and give a beautiful 
demonstration of the unification of the electromagnetic and weak forces.  The new data will be presented in detail and 
comparisons with the latest predictions of the Standard Model shown.  The H1 and ZEUS collaborations have also performed 
fits of the parton distribution functions in the proton; the results of these fits will also be presented. }

\section{Introduction} 

In this paper, measurements of deep inelastic $ep$ scattering (DIS), at high virtualities, $Q^2$, of the exchanged boson, at HERA 
are presented.  Both the H1 and ZEUS collaborations have recently published inclusive measurements for the full HERA data set 
of neutral and charge current DIS at high $Q^2$ with incoming electrons or positrons of positive or negative polarisation.  
The neutral current reaction is mediated by a photon or a $Z^0$ boson, dependent on the $Q^2$ value, and the charge current 
reaction is mediated by $W^\pm$ bosons.  The data 
provide a probe of the electroweak structure of the Standard Model and in particular verify the unification of electromagnetism and 
the weak force at scales around the masses of the weak bosons and test the chiral nature of the electroweak force.  The data also 
provide strong constraints on the structure of the proton allowing precise determinations of the parton density functions (PDFs) of 
the proton.  This elucidates one of the most important questions in science, the structure of matter at its most fundamental level, and 
on a practical level provides crucial input to other facilities colliding protons such as the LHC.  A final motivation for these measurements 
is that they probe the highest ever scales for $ep$ scattering and so may elucidate new physical phenomena, such as quark 
substructure, or provide strong constraints on their existence.

\begin{figure}[htp]
\begin{center}
~\includegraphics[trim = 0cm 4cm 0cm 0cm, clip, height=9cm]{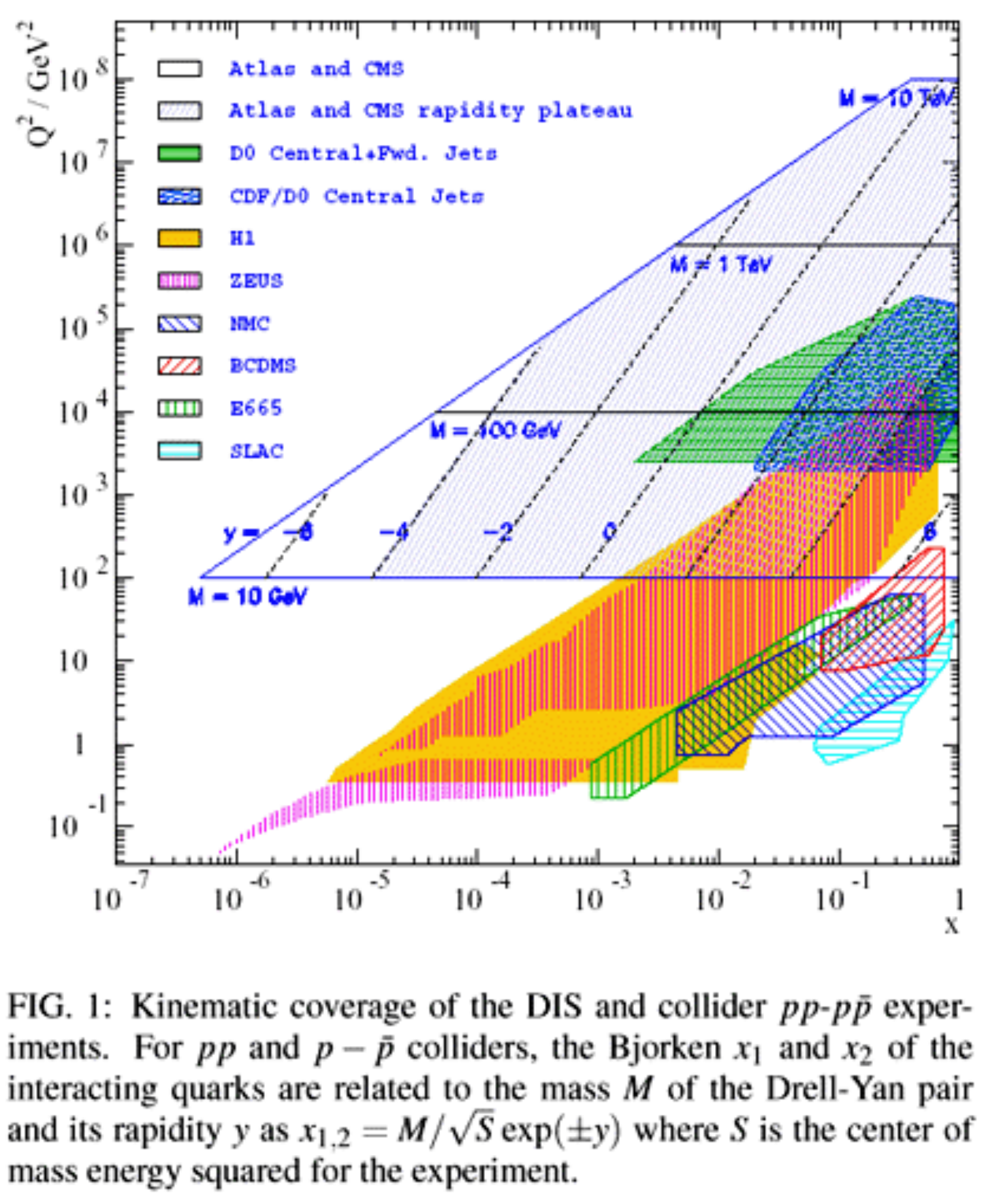}
\caption{Kinematic coverage for DIS experiments and $pp / p\bar{p}$ colliders.}
\label{fig:kin}
\end{center}
\end{figure}

The kinematic coverage of the HERA results, in $Q^2$ and $x$, the fraction of the proton's momentum carried by the struck parton,  
is shown compared to fixed-target DIS experiments and $pp / p\bar{p}$ colliders in Fig.~\ref{fig:kin}.  This demonstrates that the HERA 
data overlap with both the low-energy DIS data and the high-energy hadron collider region.  The results in these proceedings focus on 
the region above about $Q^2 \sim 100$\,GeV$^2$ and so these data will constrain the parton densities in the region of overlap between 
HERA and LHC data.  As the QCD evolution in $Q^2$ is known, this then allows predictions of the parton densities at the much higher 
LHC energies to be made where there is no HERA data, but using the precise data shown here as input.

The unpolarised neutral current cross section can be written as 

\begin{equation}
\frac{d^2\sigma_{\rm NC}^{e^\pm p}}{dx dQ^2} = \frac{2 \pi \alpha^2}{x Q^4} \left[ Y_+ F_2 \mp Y_- x F_3 - y^2 F_L  \right]
\label{eq:nc}
\end{equation}
where $\alpha$ is the fine structure constant, $y$ is the inelasticity and is related to the other DIS variables by $y = Q^2/(x s)$, where 
$s$ is the squared centre-of-mass energy and $Y_\pm = 1 \pm (1-y)^2$.  The structure functions represent the following\,:  $F_2$ 
is the sum of $q$ and $\bar{q}$ densities; $x F_3$ is the density of valence quarks; and $F_L$ is the gluon density which is important 
at high $y$ and low $Q^2$.

The unpolarised charge current cross section can be written as 

\begin{equation}
\frac{d^2\sigma_{\rm CC}^{e^\pm p}}{dx dQ^2} = \frac{G_F^2}{2 \pi} \frac{M_W^2}{Q^2 + M_W^2} \tilde{\sigma}^{e^\pm p}
\label{eq:cc}
\end{equation}
where $G_F$ is Fermi's constant and $M_W$ is the mass of the $W$ boson.  The reduced cross section, $\tilde{\sigma}$, is 
sensitive to $\bar{u}, \bar{c}$ and $d,s$ quarks for $e^+p$ data and to $\bar{d}, \bar{s}$ and $u,c$ quarks for $e^-p$ data and 
so allows deconvolution of the quark densities which are just summed in $F_2$.

Inclusion of lepton polarisation modifies the above cross sections leading to sensitivity of for example the parity violating structure 
function, $F_2^{\gamma Z}$, and the couplings of the quarks to the $Z$ boson.  The physics and formulae will be discussed when 
presenting the appropriate results.

\section{Experimental details} 

The HERA collider operated during 1992--2007 colliding electrons or positrons of 27.5\,GeV with protons of (usually) 920\,GeV giving 
a centre-of-mass energy of about 320\,GeV.  The period 1992--2000 is termed ``HERA I" and the period 2002--2007, after the 
HERA and detector 
upgrades, is termed ``HERA II".  Lower energy proton data was also taken during the early years of running and during the 
final period in order to measure the longitudinal proton structure function.  Both H1 and ZEUS collected integrated luminosities of 
about 0.5\,fb$^{-1}$ each with roughly equal amounts of $e^-p$ and $e^+p$ data.  The HERA II period represents about 75\% of the 
data and during this period polarised lepton beams were delivered with roughly equal amounts of negative and positive polarisation 
(of magnitude 30\%).

\begin{figure}[htp]
\centering
\begin{minipage}{.5\textwidth}
  \centering
  \includegraphics[angle=270,width=1.\linewidth]{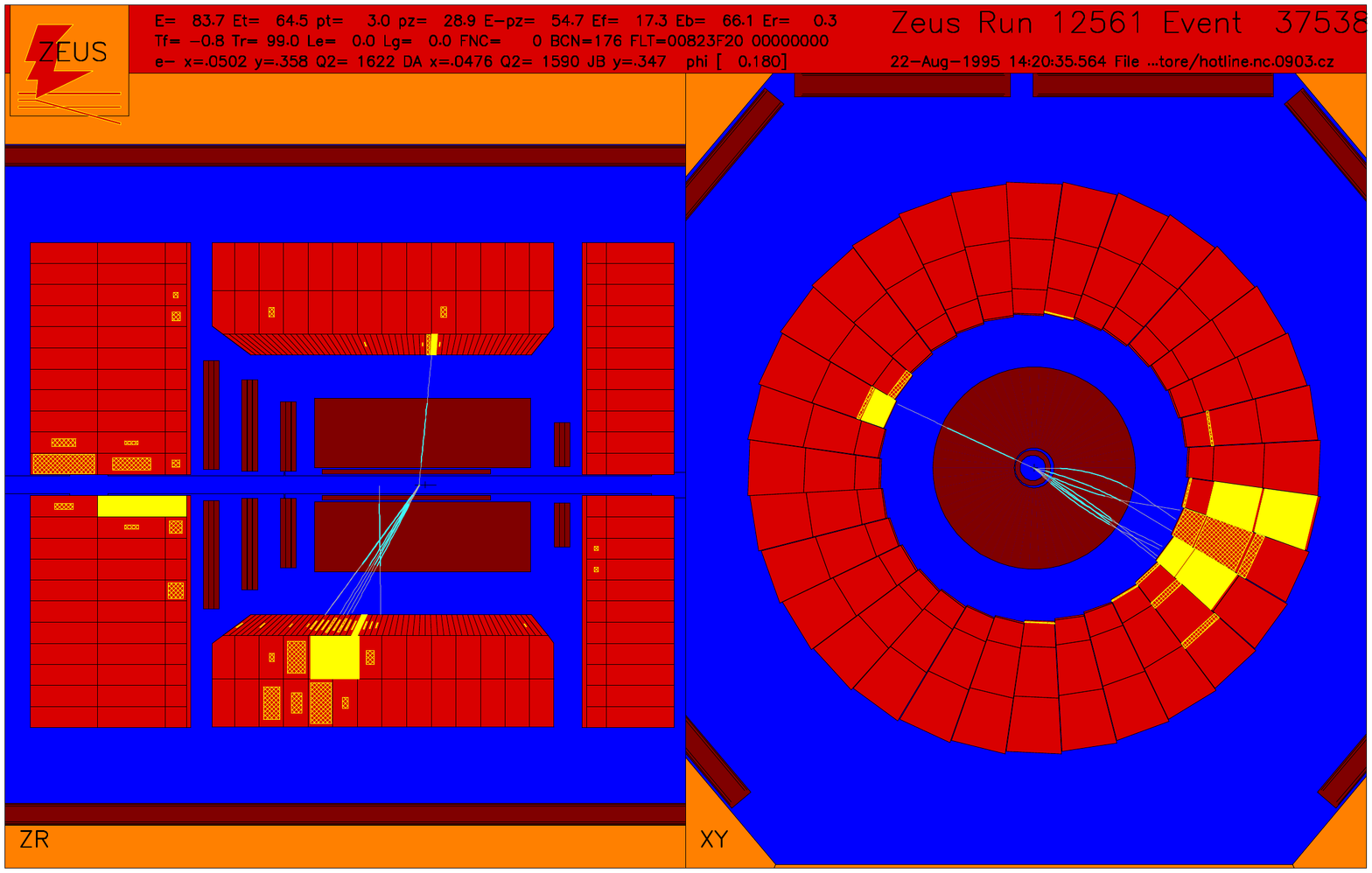}
\end{minipage}%
\begin{minipage}{.5\textwidth}
  \centering
  \includegraphics[angle=90,width=1.\linewidth]{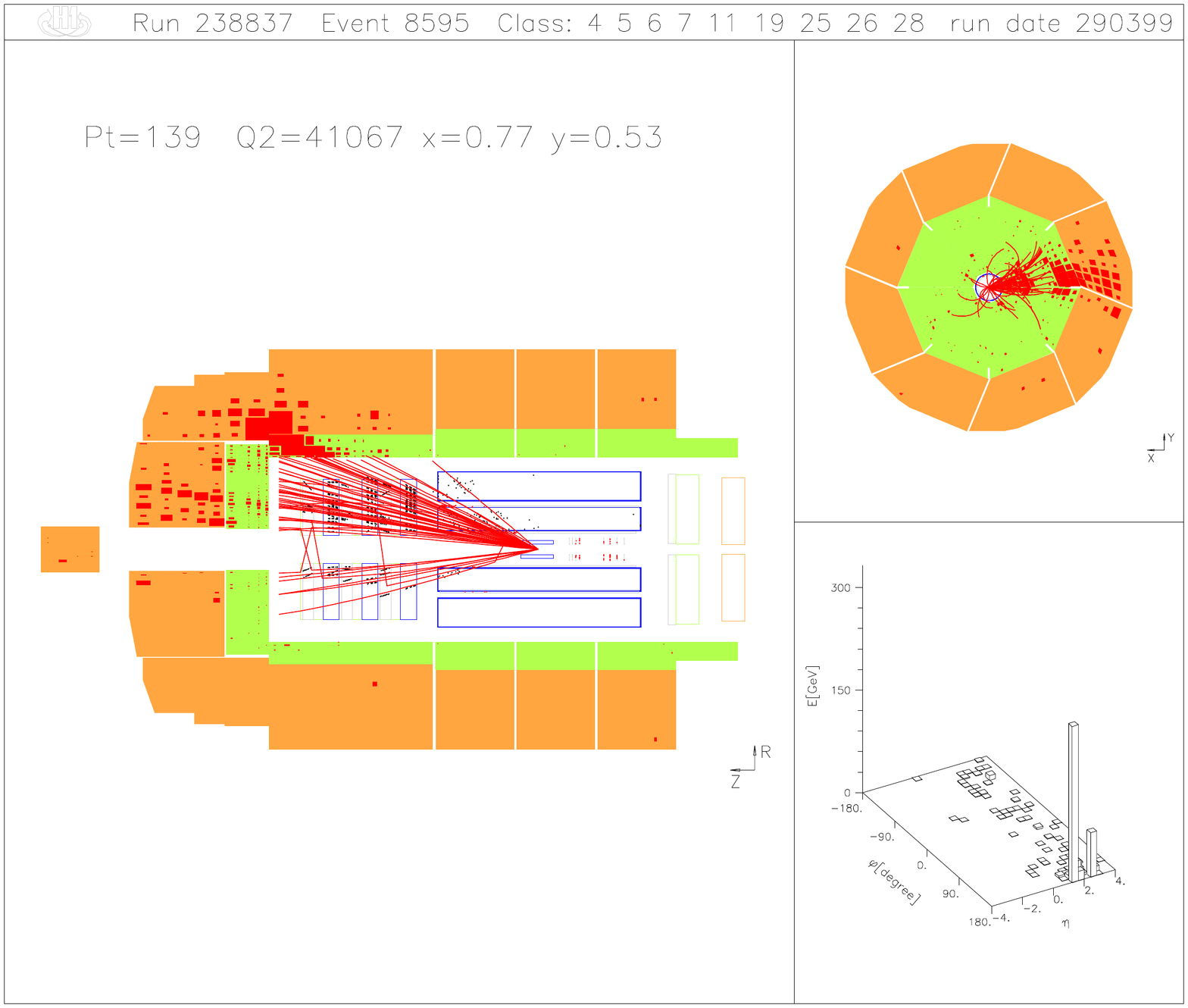}
\end{minipage}
\put(-200,80){\makebox(0,0)[tl]{\bf (a)}}
\put(-35,80){\makebox(0,0)[tl]{\bf (b)}}
\caption{Event displays showing (a) a neutral current event in the ZEUS detector and (b) a charge current event in the H1 detector.}
\label{fig:displays}
\end{figure}

Example neutral and charge current events are shown in the detector displays in Fig.~\ref{fig:displays}.  As can be seen from the figure, 
the detectors are reasonably standard for high energy colliders with vertex detectors, tracking systems, calorimeters and muon 
detectors.  In these representations, the electron enters the detector from the left and the proton enters from the right which is the reason 
for more instrumentation in the direction of the proton due to its much higher energy.  The neutral current event shown in 
Fig.~\ref{fig:displays}(a) has a clear signature of a scattered, high-energy, isolated electron back-to-back with a hadronic jet.  The charge 
current event shown in Fig.~\ref{fig:displays}(b) also has a hadronic jet but is balanced by a high-energy neutrino and so not observed 
in the detector and the missing energy is reconstructed from the kinematics.  As the high-energy electron is a clearer signal than the 
reconstructed missing energy, the neutral current sample are generally more accurately reconstructed and have lower backgrounds 
than the charge current sample.

The results presented here focus on the completion of the analyses of the inclusive neutral and charge current data from the HERA II 
period in which data is taken with both negative and positive polarised electron and positron beams.  The H1 Collaboration 
published all results in one extensive paper~\cite{h1-pub} whereas ZEUS published separately according to lepton running and 
process; the most recent discussed here are the neutral current~\cite{z-nc-eplus} and charge current~\cite{z-cc-eplus} $e^+p$ data 
with the neutral current~\cite{z-nc-eminus} and charge current~\cite{z-cc-eminus} $e^-p$ data published earlier.

\begin{figure}[htp]
\begin{center}
~\includegraphics[trim = 0cm 0cm 15cm 22cm, clip, height=4cm]{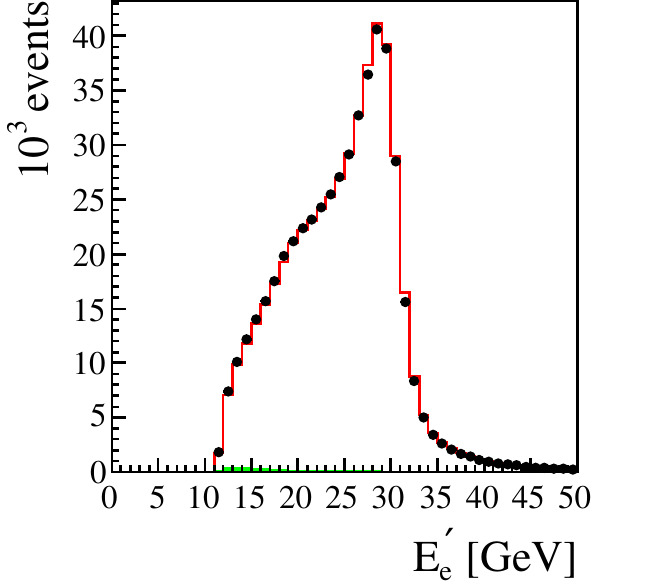}
~\includegraphics[trim = 0cm 0cm 12cm 22.5cm, clip, height=4.cm]{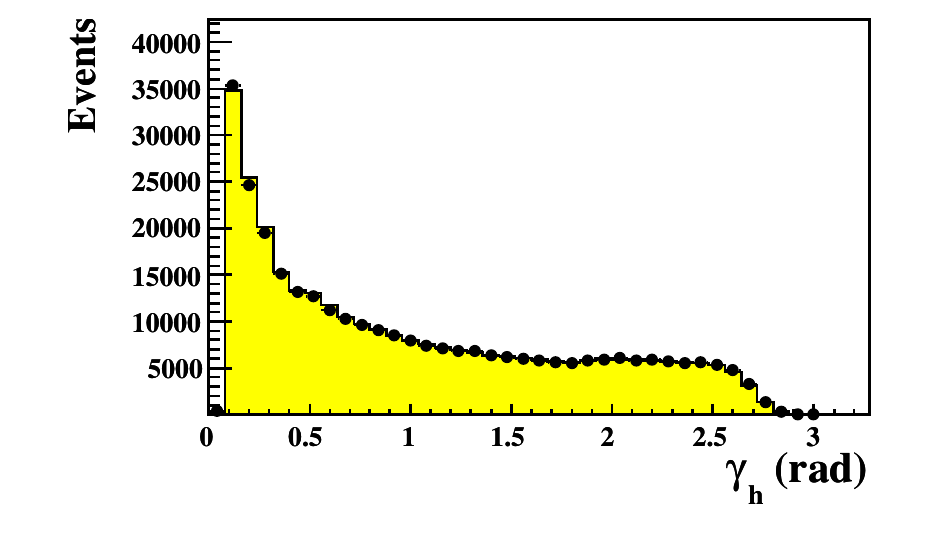}
\put(-300,100){\makebox(0,0)[tl]{\bf H1}}
\put(-100,100){\makebox(0,0)[tl]{\bf ZEUS}}
\put(-240,100){\makebox(0,0)[tl]{\bf (a)}}
\put(-35,100){\makebox(0,0)[tl]{\bf (b)}}
\caption{Example control plots for neutral current DIS events, (a) the electron energy, $E_e^\prime$, and (b) the polar angle of the 
hadronic system, $\gamma_h$, for data (points) and Monte Carlo simulations of neutral current events (histogram).  The 
background from primarily photoproduction is not visible on these scales.}
\label{fig:nc-control}
\end{center}
\end{figure}

As stated above, neutral current events are clean events which are accurately reconstructed via a variety of methods depending 
on the exact phase space or region of the detector.  Examples of how well the data are reconstructed are shown in Fig.~\ref{fig:nc-control} 
where the scattered electron energy and angle of the hadronic system are compared to neutral current Monte Carlo simulation.  
All other variables are similarly well described and such control leads to a small systematic uncertainty of 1.5\%.

\begin{figure}[htp]
\begin{center}
~\includegraphics[trim = 0cm 0cm 12cm 19cm, clip, height=4.3cm]{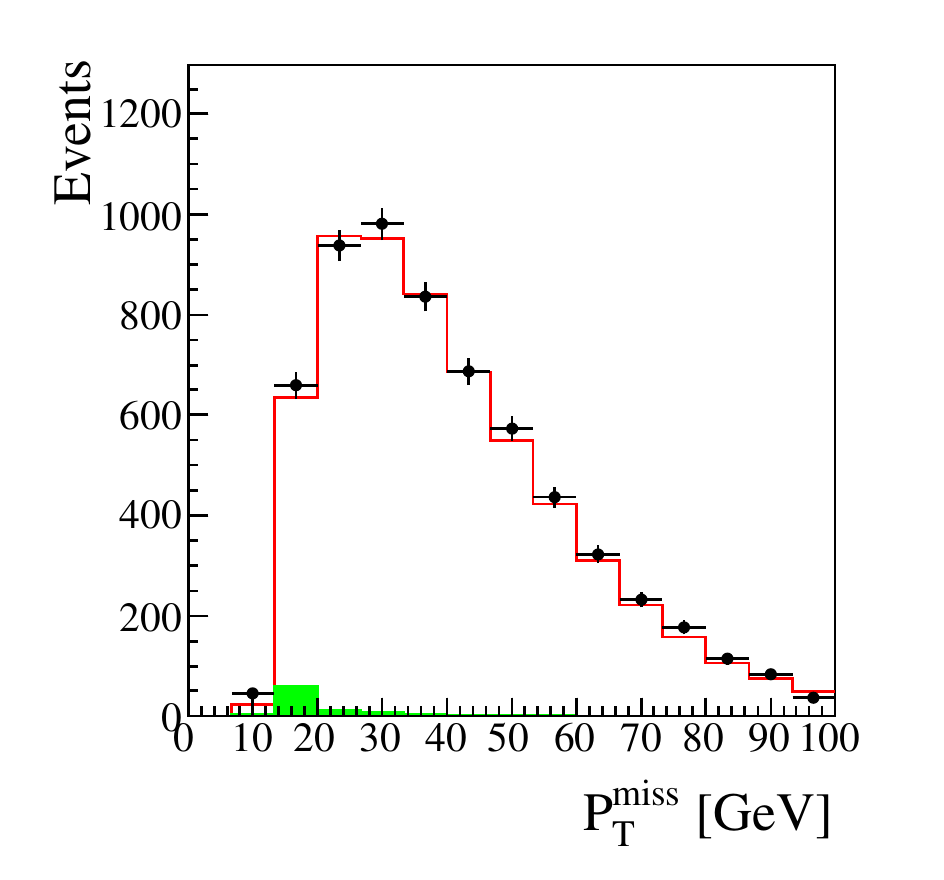}
~\includegraphics[trim = 9.5cm 15.5cm 2.5cm 4.6cm, clip, height=4cm]{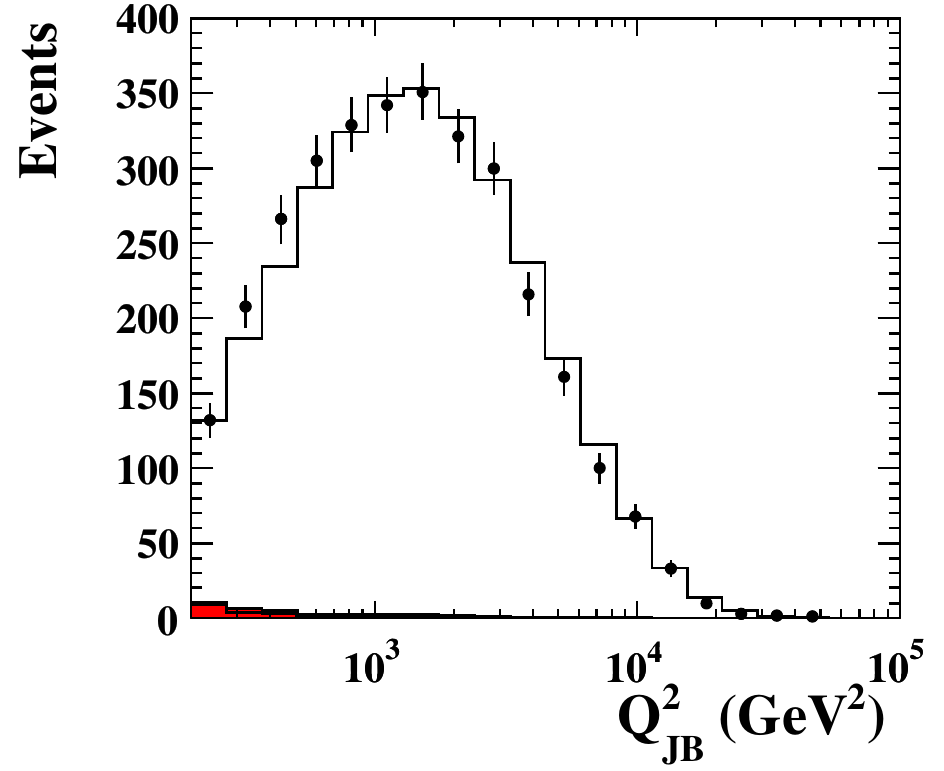}
\put(-205,105){\makebox(0,0)[tl]{\bf H1}}
\put(-65,105){\makebox(0,0)[tl]{\bf ZEUS}}
\put(-180,105){\makebox(0,0)[tl]{\bf (a)}}
\put(-27,105){\makebox(0,0)[tl]{\bf (b)}}
\caption{Example control plots for charge current DIS events, (a) the missing transverse momentum, $P_T^{\rm miss}$, and (b) the 
exchanged boson virtuality reconstructed using the Jacquet--Blondel method, $Q^2_{\rm JB}$, for data (points) and Monte 
Carlo simulations of charge current events (histogram).  The background from primarily photoproduction is visible at low 
$P_T^{\rm miss}$ and low $Q^2_{\rm JB}$.}
\label{fig:cc-control}
\end{center}
\end{figure}

In charge current events, the hadronic system and missing energy need to be well understood; the comparison of the  
Monte Carlo simulation with the data is shown in Fig.~\ref{fig:cc-control}.  Due to the missing energy, the description by the Monte Carlo 
is not as good as in neutral current events, but still under control.  The background is also somewhat larger, but still small.  
The final systematic uncertainty is about 3\%.

\section{High-{\boldmath $Q^2$} electroweak physics}

Measuring the neutral current and charge current cross sections over a wide range of $Q^2$ and in particular at values up to 
$M_{W,Z}^2$ allows the relationship to be investigated between the electromagnetic and weak interactions and, particularly due 
to using polarised electron and positron beams, understand the chiral structure of the electroweak interaction.

\begin{figure}[htp]
\begin{center}
~\includegraphics[trim = 0cm 0cm 2cm 9cm, clip, height=9cm]{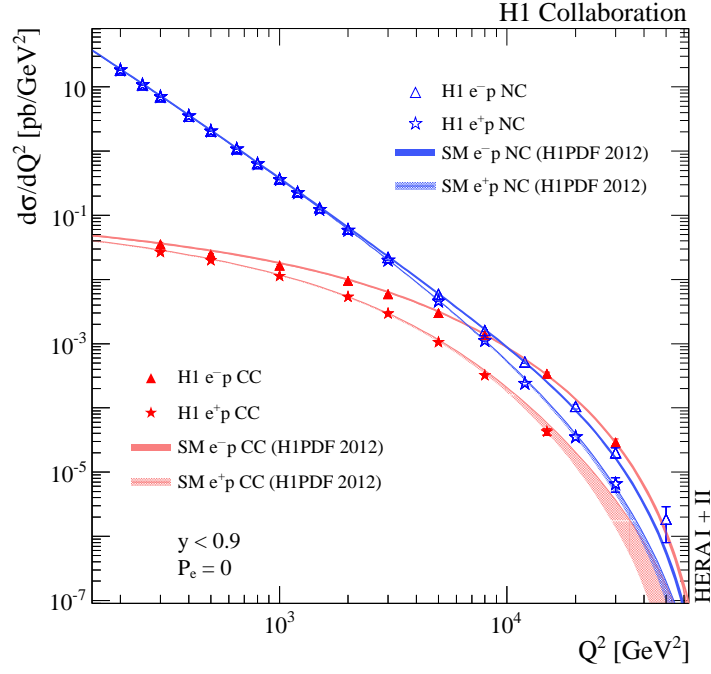}
\caption{Differential cross-section $d\sigma/dQ^2$ versus $Q^2$ for neutral (open points) and charge (closed points) current 
data for the unpolarised $e^-p$ (triangles) and $e^+p$ (stars) data compared to Standard Model predictions.}
\label{fig:nc-cc}
\end{center}
\end{figure}

The $Q^2$ dependence of the neutral and charge current unpolarised cross sections is shown in Fig.~\ref{fig:nc-cc}.  At 
$Q^2 \approx 200$\,GeV$^2$, the neutral current cross section is over two orders of magnitude higher than the charge current 
cross section.  Given the dependencies of the cross sections in Eqs.~\ref{eq:nc} and~\ref{eq:cc}, the neutral current cross 
section initially falls much faster with increasing $Q^2$ due to the dominance of photon exchange.  The cross sections become 
of similar size and have a similar $Q^2$ dependency at about $10^4$\,GeV$^2$.  The equivalence of the photon and $Z^0$ 
exchange in the neutral current process and the $W^\pm$ exchange in the charge current process illustrate the unified 
behaviour of the electromagnetic and weak interactions.

\begin{figure}[htp]
\begin{center}
~\includegraphics[trim = 0cm 0cm 3cm 15cm, clip, height=7cm]{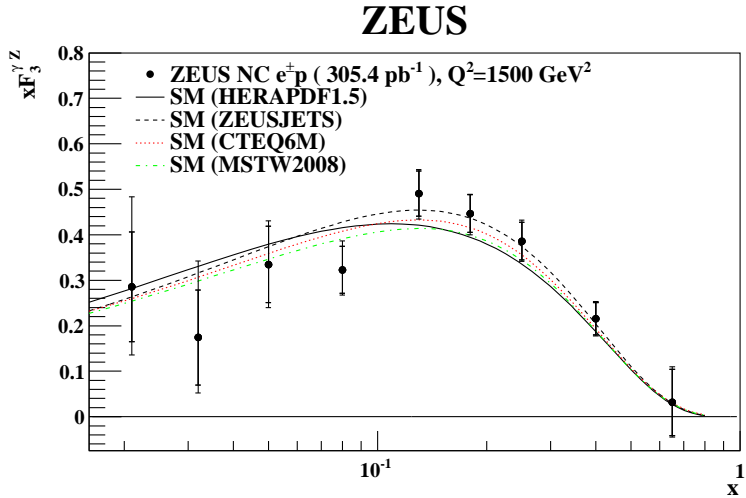}
\caption{Structure-function $xF_3^{\gamma Z}$ transformed to $Q^2 = 1\,500$\,GeV$^2$ for data (points) and the Standard 
Model expectations from various sets of PDFs.}
\label{fig:f3-gammaZ}
\end{center}
\end{figure}

The difference in the $e^+p$ and $e^-p$ unpolarised neutral current cross sections gives a direct measure of the 
structure-function $xF_3^{\gamma Z}$.  Differences in the cross sections start for $Q^2 > 1\,000$\,GeV$^2$ and increase 
with increasing $Q^2$.  The extracted $xF_3^{\gamma Z}$ is shown in Fig.~\ref{fig:f3-gammaZ} transformed to one value 
of $Q^2$; the data are well described by predictions of the Standard Model using different proton PDFs.

\begin{figure}[htp]
\begin{center}
~\includegraphics[trim = 0cm 0cm 2cm 9cm, clip, height=9cm]{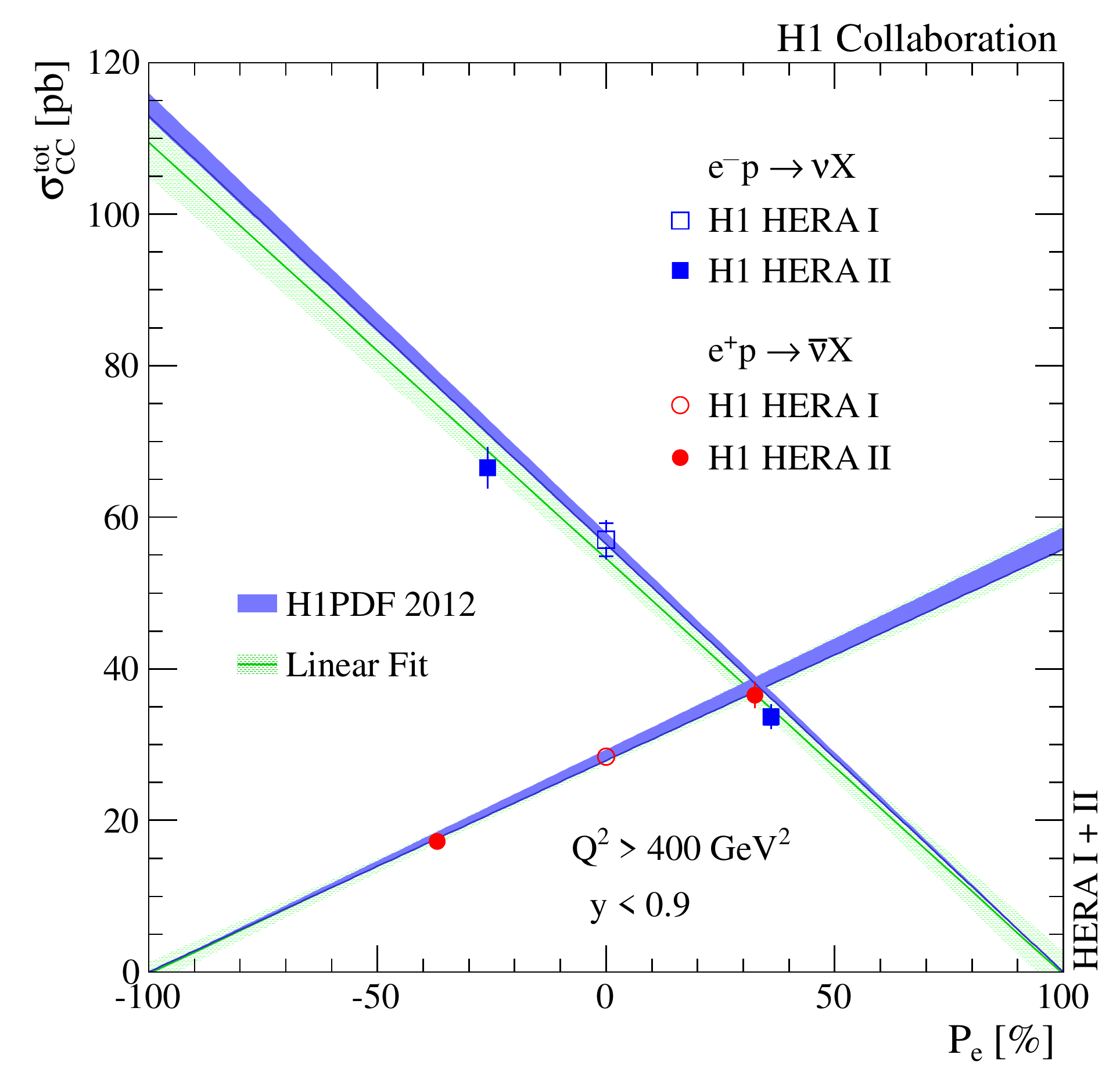}
\caption{Dependence of the charge current cross section on polarisation of the electron-beam (squares) and positron-beam 
(circles) data.  Predictions from the Standard Model (dark shaded band) and a 
linear fit to the data with its one standard deviation contour (line and light shaded band) are also shown.}
\label{fig:cc-tot}
\end{center}
\end{figure}

Investigating the cross-section dependence on the lepton beam charge and polarisation reveals, in particular for the charge 
current process, a rich chiral structure.  Figure~\ref{fig:cc-tot} shows the strong dependence of the total charge current cross 
section on the lepton-beam polarisation and charge which is well described by Standard Model predictions.  A linear fit to 
the data shows that the data are consistent with zero cross section when the positron beam has 100\% negative polarisation 
and the electron beam has 100\% positive polarisation.  The results exclude the existence of charge current events involving 
right-handed fermions mediated by a boson of mass $M_W^R$ below about 200\,GeV assuming Standard Model 
couplings and a light right-handed $\nu_e$.

\begin{figure}[htp]
\begin{center}
~\includegraphics[trim = 0cm 0cm 2cm 15cm, clip, height=7cm]{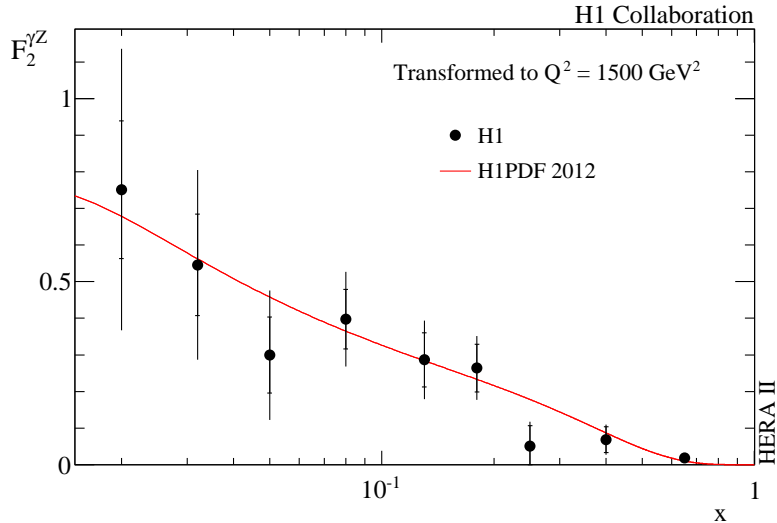}
\caption{Structure-function $F_2^{\gamma Z}$ transformed to $Q^2 = 1\,500$\,GeV$^2$ for data (points) and the Standard 
Model expectation (line).}
\label{fig:f2-gammaZ}
\end{center}
\end{figure}

Although weaker than in charge current processes, the Standard Model also predicts a difference in the neutral current cross 
section for leptons of different helicity.  A difference in the cross sections for negative and positive lepton polarisations has been 
observed in the data, starting at about $Q^2 > 1\,000$\,GeV$^2$ and increasing with increasing $Q^2$.  The Standard 
Model predictions describe the data well and therefore the data confirm parity violation effects of the electroweak interaction.  
Using these high-$Q^2$ polarised neutral current data, the parity violating structure-function $F_2^{\gamma Z}$ can be 
extracted for the first time.  The difference of the $e^+p$ and $e^-p$ cross sections, 
$(\sigma^\pm(P_L^\pm) - \sigma^\pm(P_R^\pm))/(P_L^\pm - P_R^\pm)$, where $P_L^\pm$ and $P_R^\pm$ are the negative 
and positive polarisations for $e^\pm p$ data leads to the cancellation of the $x F_3^Z$ and $x F_3^{\gamma Z}$ terms and 
allows the direct extraction of $F_2^{\gamma Z}$~\cite{h1-pub}.  The result is shown in Fig.~\ref{fig:f2-gammaZ} and is well described by 
theory.

\begin{figure}[htp]
\begin{center}
~\includegraphics[trim = 0cm 0cm 3cm 3.5cm, clip, height=12cm]{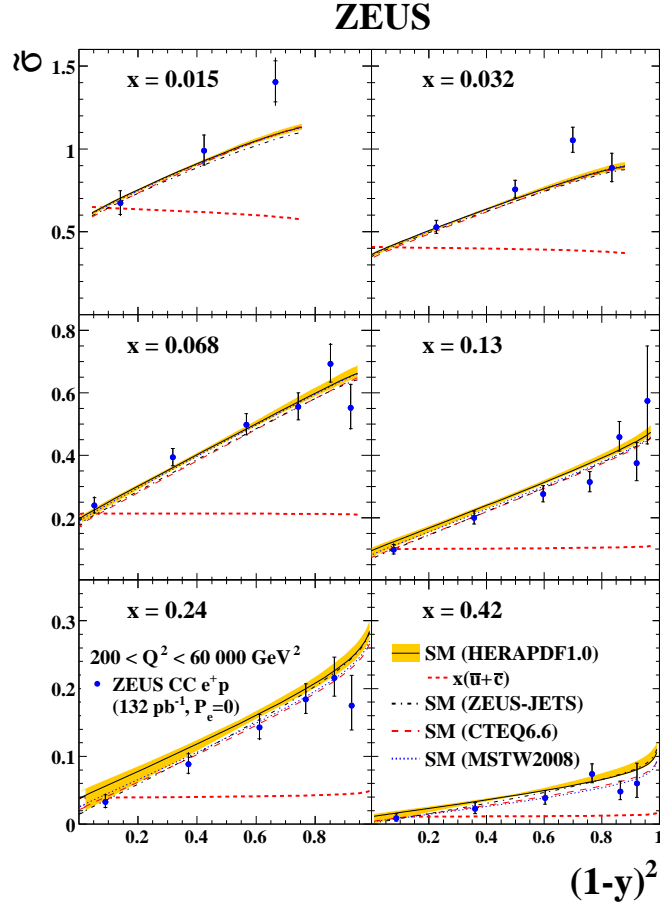}
\caption{The $e^+p$ charge current reduced cross section versus $(1-y)^2$ for fixed $x$ for data (points) and Standard 
Model predictions using different proton PDFs.  The dashed lines show the PDF contribution $x(\bar{u} + \bar{c})$.}
\label{fig:cc-diff}
\end{center}
\end{figure}

Deconvolution of the different contributions to the charge current cross section can be seen in the reduced cross sections 
versus $x$ at fixed $Q^2$ and showing the Standard Model predictions separately for the quark flavour, e.g. 
$x(\bar{u}+\bar{c})$~\cite{h1-pub,z-cc-eplus}.  This can also be seen more strikingly by considering the dependence on the 
angular distribution of the scattered quark\,: in $e^+\bar{q}$ charged current DIS, the angular distribution will be flat in the 
positron-quark centre-of-mass scattering angle, $\theta^*$; whereas it will exhibit a $(1 + \cos\theta^*)^2$ distribution in 
$e^+q$ scattering.  Since $(1-y)^2 \propto (1 + \cos\theta^*)^2$, the reduced cross section versus $(1-y)^2$ is shown, for 
fixed $x$, in Fig.~\ref{fig:cc-diff}.  The data are well described by the Standard Model predictions.  At leading order in QCD, 
the intercept of the prediction gives the $(\bar{u} + \bar{c})$ contribution, while the slope gives the $(d+s)$ contribution.

\begin{figure}[htp]
\begin{center}
~\includegraphics[trim = 3.2cm 19cm 3.2cm 1cm, clip, height=6cm]{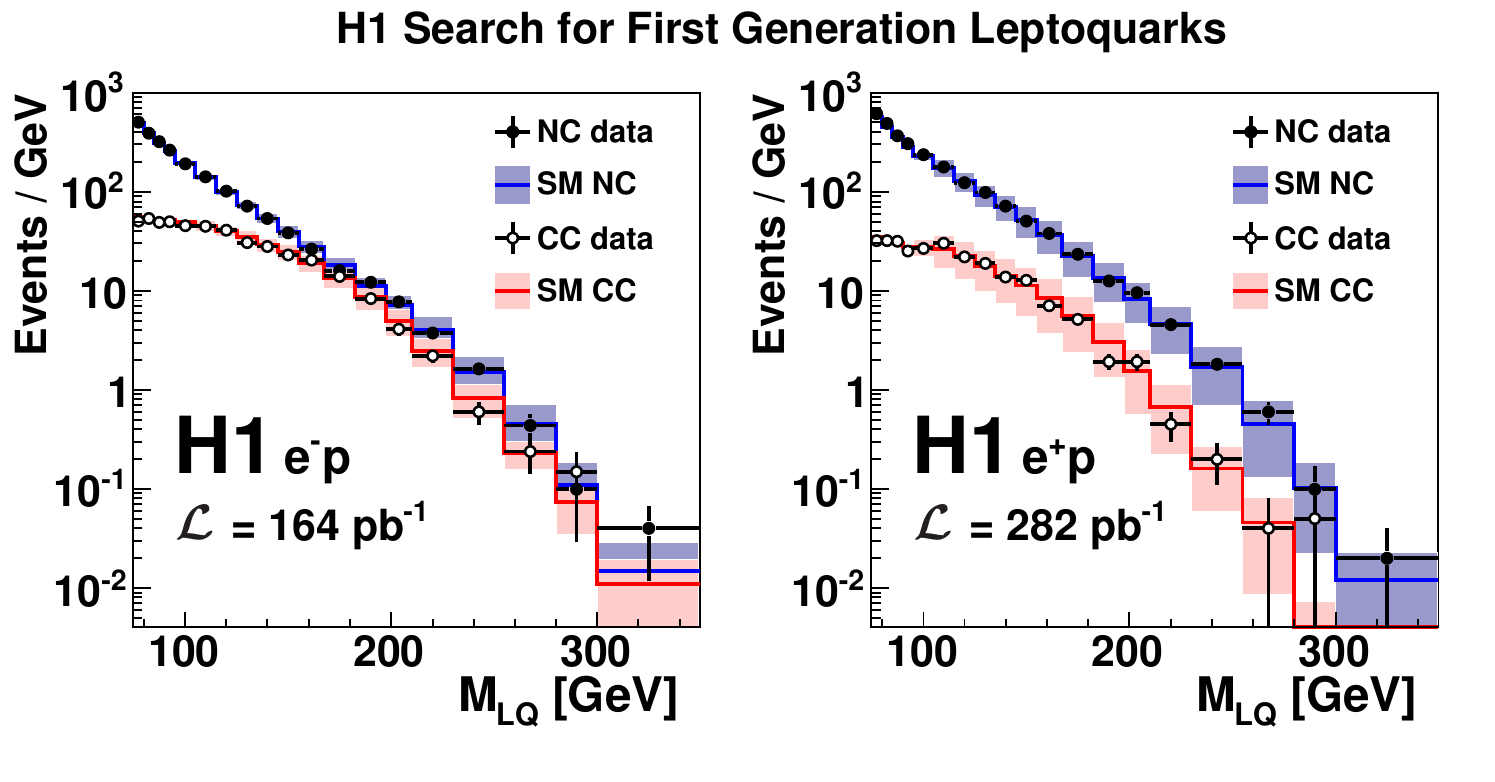}
~\includegraphics[trim = 0cm 0cm 3cm 14cm, clip, height=6cm]{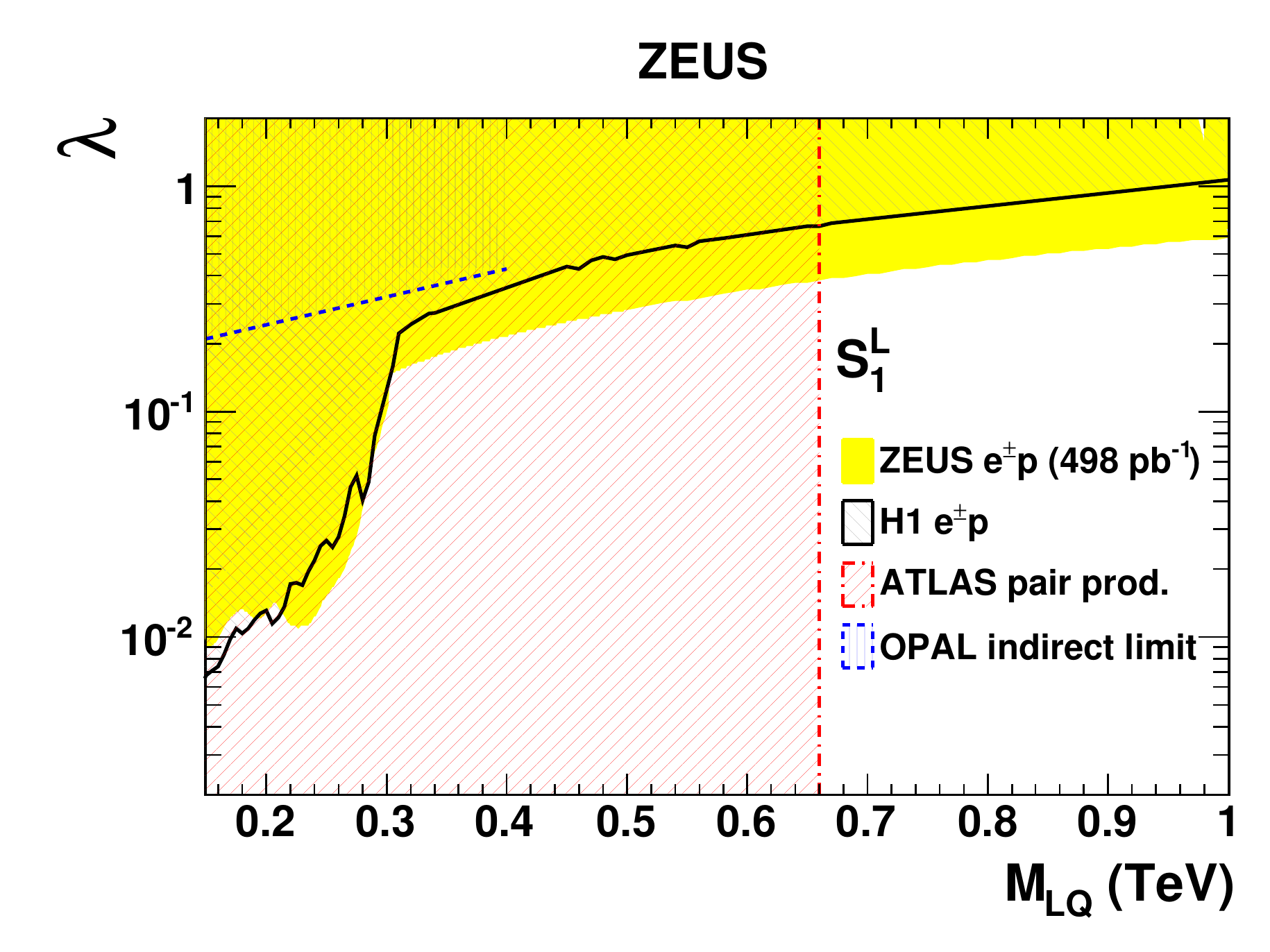}
\caption{The neutral and charge current $e^-p$ and $e^+p$ data as a function of the leptoquark mass, 
$M_{\rm LQ} = \sqrt{Q^2/y}$, compared to Standard Model predictions.  Limits on the Yukawa coupling, 
$\lambda$, versus $M_{\rm LQ}$ showing HERA data compared to ATLAS $pp$ and OPAL $e^+e^-$ data.}
\label{fig:lq}
\end{center}
\end{figure}

As inclusive measurements of the DIS cross section provide data at the highest scales, they can be used to search for 
new high-energy phenomena such as leptoquarks~\cite{h1-lq,zeus-lq} or quark substructure~\cite{h1-ci}.  Given the 
incoming electron beam and quark 
from the proton, HERA provides a unique opportunity to search for resonantly-produced leptoquarks.  Recently, both 
collaboration have published results on searches for leptoquarks in neutral and charge current events using the full 
data sample.  An example ``leptoquark" mass spectra is shown in Fig.~\ref{fig:lq} and compared to 
Standard Model predictions where a resonant-mass structure would be expected for leptoquark production.  Given that 
neither experiment observed an significant deviation from the Standard Model, limits on the production of leptoquarks 
were extracted.  Example results are presented in Fig.~\ref{fig:lq} in terms of limits on the Yukawa coupling, $\lambda$, 
as a function of $M_{\rm LQ}$ for a particular type of leptoquark  ($S_1^L$).  The limits extend significantly beyond those 
from $e^+e^-$ experiments and for high values of $\lambda$ also extend up to higher masses than the results from the 
LHC where the results are independent of the coupling.

\section{Comparison to theory and extracted PDFs}

Throughout the previous section the focus was on what can be learnt from considering the general theory of the Standard 
Model, although it was noted that the specific predictions incorporating electroweak and QCD effects describe the data well.  
In this section, the description of specific predictions is discussed and presented in more detail, in particular QCD fits to the 
data and the improvement in the proton PDFs using these data are presented.

\begin{figure}[htp]
\begin{center}
~\includegraphics[trim = 0cm 0cm 2cm 9cm, clip, height=9cm]{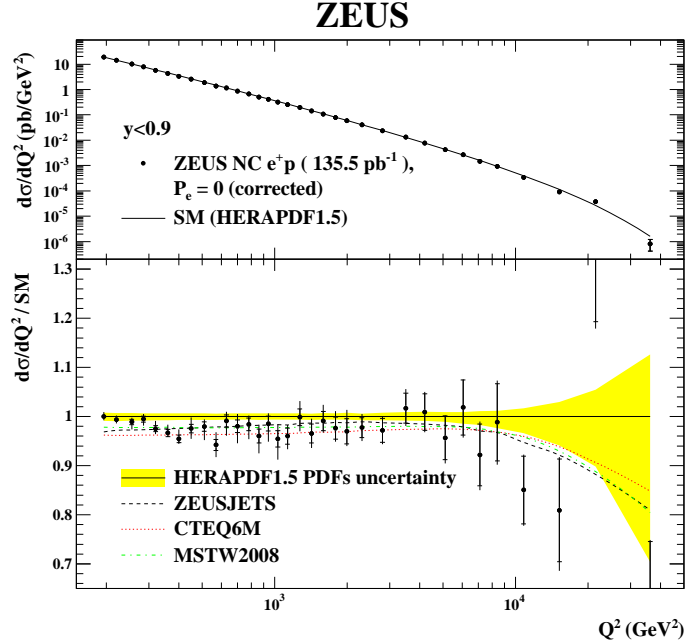}
\caption{Neutral current cross-section $d\sigma/dQ^2$ for unpolarised $e^+p$ data (points) compared to the Standard Model 
with different choices of proton PDFs.  The lower half shows the ratio of the data and other predictions to the Standard Model 
using the HERAPDF1.5.}
\label{fig:nc-theory}
\end{center}
\end{figure}

The $e^+p$ neutral current cross-section $d\sigma/dQ^2$ is shown in Fig.~\ref{fig:nc-theory} compared with Standard 
Model predictions using different protons PDFs, HERAPDF1.5~\cite{herapdf}, ZEUSJETS~\cite{zeusjets}, 
CTEQ6M~\cite{cteq} and MSTW2008~\cite{mstw}.  The HERAPDF1.5, ZEUSJETS and H1PDF2012~\cite{h1-pub} (to be 
discussed in more detail) are extracted using their own respective data with a complete understanding of the correlations 
of the uncertainties.  The CTEQ6M and MSTW2008 are global fits incorporating many additional different data sets such 
as from fixed-target DIS experiments, jet production at hadron colliders, etc..  These PDF fits used previous HERA data 
but not that shown in Fig.~\ref{fig:nc-theory}; the HERAPDF1.5 used all other ZEUS data and preliminary versions of the 
HERA II high-$Q^2$ data from H1.  The predictions describe the data well over the seven orders of magnitude 
in the cross section.  The charge current data is similarly well described and the new data will be able to constrain the 
PDFs further.

The H1 Collaboration have done an NLO QCD fit to all their DIS data, H1PDF2012, also repeating the fit without the HERA II data to 
demonstrate the effect of the new data.  The fit strategy follows closely that adopted for the determination of the HERAPDF1.0 
sets and the reader is referred there~\cite{herapdf1} as well as the H1 publication~\cite{h1-pub} for more details.  Simply put, 
the procedure relies on the factorisation of the DIS cross section into the short-distance cross section calculable in pQCD and 
the proton PDFs evolved with the $Q^2$ by the DGLAP equations which are fit to the data.  The fits are done in NLO QCD and 
also now NNLO QCD with a scale chosen as $Q$ and assumptions such as the starting scale, the functional form of the 
individual PDFs, the heavy quark masses, etc.\ which are then varied to assess the uncertainty on the procedure.  The 
uncertainties consist of\,:  the experimental which are derived from a change in $\chi^2$ of 1 having taken into account 
correlations between data points and their uncertainties; model uncertainties which includes variations 
of the heavy-quark masses, minimum $Q^2$ and strange-quark distribution; parameter uncertainties coming from an envelope 
corresponding to the set of fits in the $\chi^2$ optimisation and from varying the starting scale.

\begin{figure}[htp]
\begin{center}
~\includegraphics[trim = 0cm 0cm 3cm 10cm, clip, height=8cm]{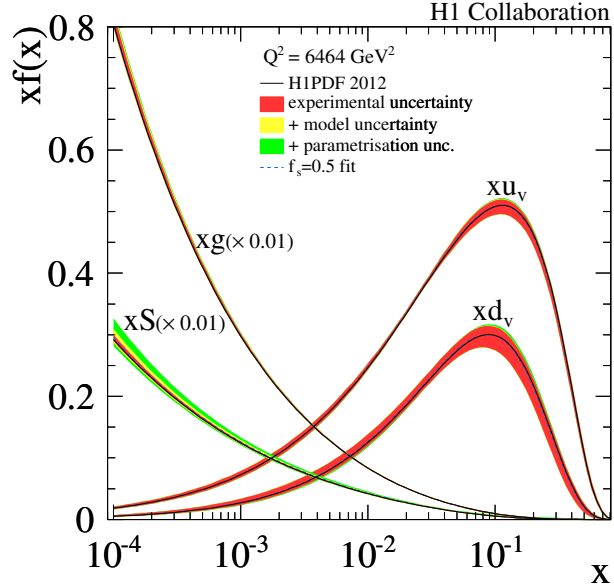}
\caption{Parton distribution functions of H1PDF2012 at the scale of $Q^2 = M_W^2 = 6464$\,GeV$^2$.  The gluon and sea 
distributions are scaled by a factor of 0.01.  The PDFs with the strange quark fraction $f_s = 0.5$ as found by 
ATLAS~\protect\cite{atlas-fs} are also shown.}
\label{fig:pdf}
\end{center}
\end{figure}

The resulting $\chi^2/{\rm ndf}$ is about 1.  Example PDFs and their uncertainties are shown evolved to $Q^2 = M_W^2$ in 
Fig.~\ref{fig:pdf}.  The valence quarks peak at about $x = 0.1$  and dominate for high values whereas the gluon and 
sea-quark densities rise quickly with decreasing $x$ and are over two orders of magnitude higher than the valence quark 
densities for $x < 10^{-3}$.  Fitting different data sets separately or combined produced results in good agreement 
demonstrating the consistency of the data samples and fit procedure.

\begin{figure}[htp]
\begin{center}
~\includegraphics[trim = 0cm 0cm 2cm 2cm, clip, height=14cm]{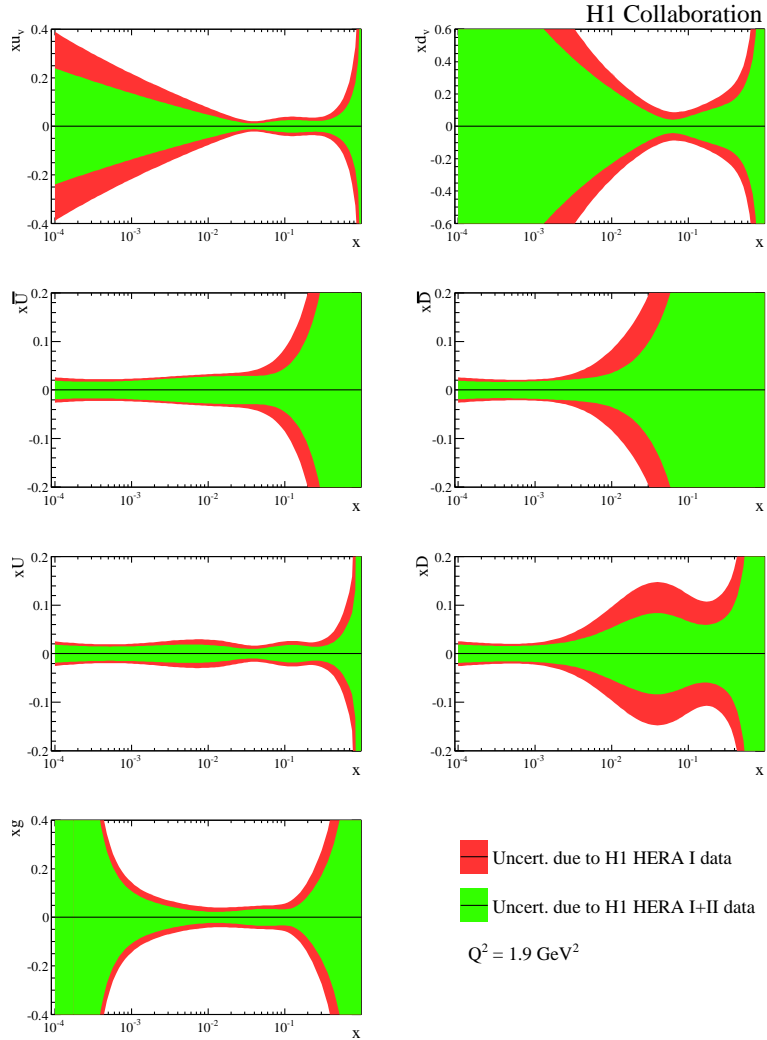}
\caption{Comparison of relative experimental uncertainties of the PDFs extracted from HERA~I (outer) and HERA II (inner) 
data under the same fit conditions.}
\label{fig:pdf-unc}
\end{center}
\end{figure}

The relative uncertainties for the PDFs are shown in Fig.~\ref{fig:pdf-unc} for the combined fit and the fit to the HERA I 
data only.  Inclusion of the HERA II data has an impact on all distributions, reducing the uncertainty by up to a factor of 
two, and particularly for the down-type quark distribution, $xD$.  The impact of the HERA~II data will also feed into any 
future HERAPDF version and global QCD fits.

\section{HERAPDFs at the LHC}

\begin{figure}[htp]
\begin{center}
~\includegraphics[trim = 0.5cm 0cm 1cm 0cm, clip, height=9cm]{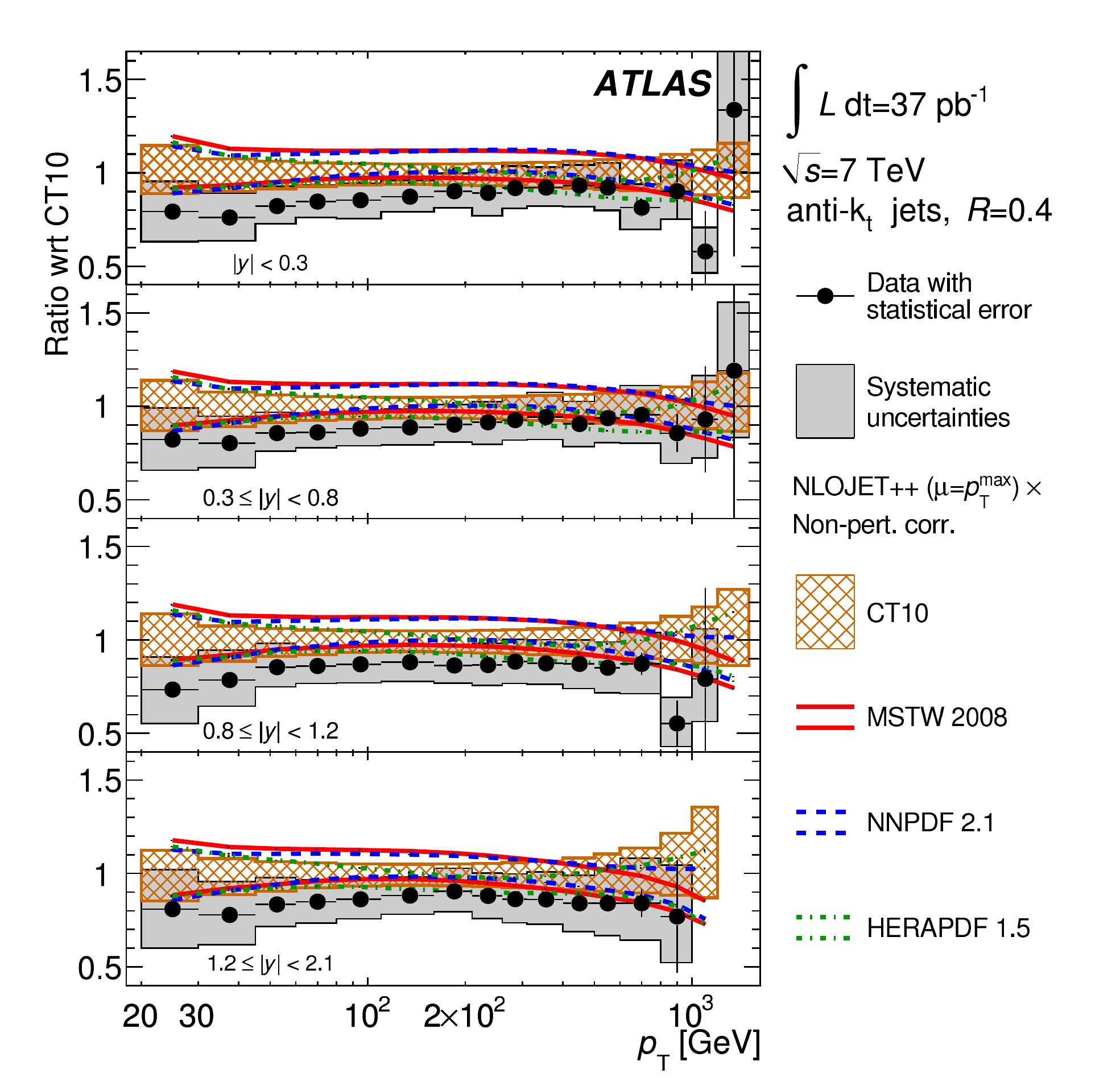}
\caption{Ratios of inclusive jet double-differential, in transverse momentum and rapidity, cross sections to the CT10 PDF 
set~\protect\cite{ct10} for ATLAS $pp$ jet data (points) and different PDF sets.}
\label{fig:atlas}
\end{center}
\end{figure}

The HERAPDFs and global PDF fits are used extensively at the LHC such as in predictions of jet production, 
e.g.~\cite{atlas-jets}.  As shown in Fig.~\ref{fig:atlas}, all the various PDF fits give a similarly good description of the 
LHC data up to jet transverse energies of 1\,TeV.  That jet production at such high scales is so well and precisely 
predicted by the different PDF fits is due dominantly to the impact of the HERA data in these fits; the high $Q^2$ 
data presented here will improve these predictions.

H1 and ZEUS have recently published a combined measurement of charm production in DIS~\cite{charm}.  As well 
as providing significantly improved results over any single measurement, it also has a significant impact constraining 
the charm mass and scheme used to calculate charm production in QCD fits.  This then provides far more precise 
cross sections of $W^\pm$ and $Z^0$ production at the LHC.

The HERA data is crucial to many aspects of LHC physics and fits using all the high $Q^2$ data shown here as well 
as the new charm results will be of great benefit to the LHC programme.


\section{Summary}

New, final results from both collaborations on the high-$Q^2$ neutral and charge current cross sections have been 
presented.  They provide many beautiful demonstrations of our understanding of electroweak physics such as its chiral 
structure and the unification of electromagnetism and weak interactions.  The data are precise and cover a wide kinematic 
range and along with the other HERA measurements provide the crucial constraints on the parton densities in the proton.  
In the future, the H1 and ZEUS high-$Q^2$ data will be combined and along with charm and jet data be used in  
HERAPDF fits.




\begin{thebibliography}{0}

\bibitem{h1-pub}  H1 Coll., F.D.~Aaron et al., JHEP {\bf 09} (2012) 061.

\bibitem{z-nc-eplus}  ZEUS Coll, H. Abramowicz et al., DESY-12-145.

\bibitem{z-cc-eplus}  ZEUS Coll, H. Abramowicz et al., Eur. Phys. J. {\bf C~70} (2010) 945.

\bibitem{z-nc-eminus}  ZEUS Coll, H. Abramowicz et al., Eur. Phys. J. {\bf C~63} (2009) 171.

\bibitem{z-cc-eminus}  ZEUS Coll, H. Abramowicz et al., Eur. Phys. J. {\bf C~61} (2009) 223.

\bibitem{h1-lq}  H1 Coll., F.D.~Aaron et al., Phys. Lett. {\bf B~704} (2011) 388.  

\bibitem{zeus-lq}  ZEUS Coll., H. Abramowicz et al., Phys. Rev. {\bf D~86} (2012) 012005.

\bibitem{h1-ci}  H1 Coll., F.D.~Aaron et al., Phys. Lett. {\bf B~705} (2011) 52.

\bibitem{atlas-fs}  ATLAS Coll., G. Aad et al., Phys. Rev. Lett. {\bf 109} (2012) 012001.

\bibitem{herapdf}  H1 and ZEUS Colls.,  \emph{QCD NLO analysis of inclusive data (HERAPDF1.5)},  
{\tt https://www.desy.de/h1zeus/combined\_results/herapdftable/}

\bibitem{zeusjets}  ZEUS Coll., S. Chekanov et al., Eur. Phys. J. {\bf C~42} (2005) 1.

\bibitem{cteq}  J. Pumplin et al., JHEP {\bf 0207} (2002) 012.

\bibitem{mstw}  A.D.~Martin et al., Eur. Phys. J. {\bf C~63} (2009) 189.

\bibitem{herapdf1}  H1 and ZEUS Colls., F.D.~Aaron et al., JHEP {\bf 01} (2010) 109.

\bibitem{atlas-jets}  ATLAS Coll., G. Aad et al., Phys. Rev. {\bf D~86} (2012) 014022.

\bibitem{ct10}  H.-L. Lai et al., Phys. Rev. {\bf D~82} (2010) 074024.

\bibitem{charm}  H1 and ZEUS Colls., F.D.~Aaron et al., DESY-12-172.

\end{thebibliography}
\end{document}